\documentclass[pre,reprint,nofootinbib,superscriptaddress]{revtex4-2}
\pdfoutput=1

\usepackage{amssymb, amsmath, amsmath, amsfonts, bm, tabularx,graphicx}
\usepackage[colorlinks,linkcolor=blue,urlcolor=blue,citecolor=blue]{hyperref}
\usepackage{physics}

\usepackage[english]{babel}

\usepackage{xcolor}
\usepackage{xspace}
\usepackage{ifthen}

\begin{document}

\title{Bertlmann's socks from a Viennese perspective.}

\author{Marcello {Poletti}}
\email{epomops@gmail.com}
\affiliation{San Giovanni Bianco, Italy}

 \begin{abstract}
Quantum mechanics is a theory that is as effective as it is counterintuitive. While quantum practices operate impeccably, they compel us to embrace enigmatic phenomena like the collapse of the state vector and non-locality, thereby pushing us towards untenable "hypotheses non fingo" stances. However, a century after its inception, we are presented with a promising interpretive key, intimated by Wheeler as early as 1974\cite{Wheeler}. The interpretative paradoxes of this theory might be resolved if we discern the relationship between logical undecidability and quantum undecidability. It will be demonstrated how both are intricately linked to an observer/observed relational issue, and how the idiosyncratic behaviours of quantum physics can be reconciled with the normative, following this path.
\end{abstract}

\maketitle

\section{A physics problem}

\begin{quote}
What, exactly, is bizarre, paradoxical, and incomprehensible in quantum mechanics?
\end{quote}

Almost everything. QM is replete with bizarre effects. The wave-particle duality, the tunnel effect, the superposition principle, non-commutativity, the use and necessity of complex numbers, the uncertainty principle, the probabilistic nature of the theory, non-locality, the measurement problem, and the list could go on. QM is a physical theory that prevents us from constructing a sensible physical picture of the world, despite the fact that, in the practice of "shut up and calculate"\cite{Mermin}, it works splendidly.

The debate around this topic has been heated since the inception of the theory. A hundred years later, a considerable number of thinkers have thrown in the towel, resigning themselves to an "it is what it is" attitude. However, an equally large group perseveres; how can we contemplate going beyond QM if we do not understand QM at all?

But in what sense do we "understand nothing"? We handle the mathematics of QM very adeptly and can make accurate predictions from it. What more is desired?
One of the most apparent problems is that QM inhibits our capacity to imagine the world. It meticulously describes the behavior of electrons or photons but fails to provide a picture of these entities. Indeed, the theory opposes the construction of an image, forcing us into a convoluted and unnatural narrative ("spin is somehow analogous to angular momentum, but we must not think of it as a spinning ball").

However, this irreducibility of the theory to the human categories of "object," "space," and "time" is not a novelty of Quantum Mechanics. Newtonian gravitation is, in the same way, an extraordinarily effective theory based on a magical substance, the gravitational field, which we can only represent non-mathematically as an "intangible magical fluid." In the same vein, the theory of the magnetic field and the problem of the luminiferous ether are situated. Certainly, the success of General Relativity also comes through the psychological sensation that Einstein managed, for the first time, to provide a complex but minimally magical or non-magical schema of the world. However, just a few years later, QM would arrive to restore the status quo, and it would do so disruptively.

Not only are the categories with which we represent the world inadequate for describing certain aspects of reality, but even the deeper philosophical prerequisites, such as realism and locality, are thrown into crisis. In this more radical sense, the theory becomes peculiar after the Einstein-Podolsky-Rosen paradox\cite{EPR}.

The EPR paradox opens a chasm that is, however, closed for many years by Bohr's principle of complementarity\cite{Laudisa2}. Personally, I find the principle of complementarity utterly obscure and incomprehensible, but Bohr's authority appears sufficient during those years, and in any case, no alternatives are in sight. Moreover, Von Neumann demonstrates that alternatives are not possible\cite{VonNeumann1932}, effectively putting to rest Einstein's philosophical whims and the residual group of nostalgic realists.

The debate nevertheless progresses, and in 1952, Bohm\cite{Bohm} proposed his pilot-wave theory, constructively refuting Von Neumann's proof and the principle of complementarity. Quantum Mechanics is not incompatible with assigning objective position and momentum, the hidden variables, to quantum particles. The theory is not taken seriously, even by Einstein himself, who dismisses it as "too cheap," but its effect is still disruptive as it opens up the concrete possibility of "another theory" that does not force us to renounce realism and locality.

However, Bohm's theory is non-local. Is it possible to provide a local version? John Bell works on this and produces a convincing negative response: no theory can respect the EPR criteria of locality and realism and be compatible with the results of QM.

Bell's theorem, over many years of reflection, experimentation, and assimilation, becomes a paradigm. It is entirely reasonable to state that today, a considerable majority of thinkers consider non-locality an established fact.

Contrary to the prevailing view, I believe that there is still room for local realism\footnote{Naturally, there are various studies and positions in this regard, such as superdeterminism\cite{hossenfelder} or 't Hooft local determinism\cite{hooft}.}, and that Bell's work is extraordinary not so much for its conclusions (that reality is non-local) but for its ability to quantify the strangeness of Quantum Mechanics.

In '64\cite{Bell64}, Bell managed to define in a concrete, clear, and pragmatic way a boundary between quantum strangeness and classical physics. QM is strange insofar as it violates certain inequalities. Bell's work is extraordinary (and finally recognised with the 2022 Nobel Prize) as it allows us to ignore the technical details of QM, shifting the focus from the quirks of the theory to the oddities of reality: it does not matter if QM is correct or not, if it is incomplete or not, if there is a better theory or not; if concrete experiments violate Bell's inequalities, it is reality itself that shows bizarre and nonsensical aspects. And reality, as has now been widely verified, particularly since Aspect's\cite{Aspect} work, violates Bell's inequalities.

In '81\cite{Bell81, bertlmann, bellspeakable}, in an iconic paper, Bell reduced his work to the essential by asking what difference there is between Professor Bertlmann's unmatched socks and two entangled electrons in a Stern-Gerlach type experiment. The answer is that experiments with socks yield results that adhere to the rules of classical probability, whereas experiments with electrons yield results that adhere to the rules of quantum probability, and the two results differ in that the electrons violate Bell's inequalities, producing outcomes that are impossible in classical probability and, therefore, in the usual macroscopic world.

Both in classical probability and in quantum mechanics, "probability" is a real number in the range [0,1]. In general, therefore, the "probability of event $p$" can coincide in the two cases. The two paths diverge when posing more complex questions, involving more measurements, such as determining the probability of $p \land q$.
In his 1981 work, Bell revisits his reasoning using an extraordinarily elementary form of inequalities, the Wigner-D'Espagnat inequality. We will consider a variant of this inequality in set-theoretic terms:
\begin{equation}\label{we}
	(A,B)\cup(\neg B,C)\supseteq(A,C)
\end{equation}
Things that are \textit{A and B} along with things that are \textit{not B and C} include things that are \textit{A and C}.

The demonstration of \ref{we} is trivial. If an element $x \in (A,C)$, then either $x \in B$ in which case $x \in (A,B)$, or $x \in \neg B$, in which case $x \in (\neg B, C)$. That's all there is to it. And yet, QM violates this banal inequality in a Stern-Gerlach type experiment, how is that possible?

What is bewildering is the fact that \ref{we} is not about QM, it's not even about physics, but is a fundamental property of set theory, the most fundamental layer of mathematics itself\cite{principiamathematica}, that mathematics in direct contact with formal logic. Does QM violate logic? Does reality violate logic? The very existence of a term such as "quantum logic"\cite{NeuBir} seems to suggest something of the sort, but this would trigger an explosive paradox. If classical logic were to be rejected, then the entire mathematical apparatus upon which Quantum Mechanics (QM) is built would also have to be discarded, as it is entirely based on those premises, and thus QM itself would have to be rejected. The proposition "QM gives rise to an alternative logic" has the flavor of a contradiction.

Alternatively, one could postulate the existence of two worlds, one classical subject to classical logic and one quantum subject to quantum logic, separated by an interface, a Heisenberg cut. This solution conflicts both with experimental evidence, which does not indicate any presence of such an interface, and with the philosophical and mathematical problems posed by having "two logics".

Bell, from his inequalities, draws philosophical conclusions, summarizing to the extreme: If quantum objects only possessed well-defined properties, like the color of socks, then these could be categorized into classical sets and would therefore respect classical probability theory. Consequently, the conjugate properties that violate Bell's inequalities cannot be well-defined a priori. As a result, the correlated behavior of entangled objects, even over great distances, necessarily entails a form of communication at a distance. Quantum Mechanics, or any theory compatible with the results of Aspect, is therefore non-local.

The necessity of non-locality is a bewildering paradox within the paradox. The mechanism by which entangled particles should "communicate at a distance" is entirely absent in QM, it is a magical phenomenon on which it is not even possible to imagine making concrete hypotheses.

However, Bell's philosophical conclusions are certainly weaker than his technical conclusions. There is a continuous oscillation, a lexical ambiguity, between what a quantum object is and what a sentient agent can know about such an object. This ambiguity constantly pervades the entire debate on the foundations\cite{Laudisa}, from its origins to current textbooks.

Does a quantum object inherently possess indefinite properties before measurement, or is it that an observer cannot assign such properties in a well-defined manner? Does the uncertainty principle specify the limits of the properties of a quantum object itself, or the limits of what we can measure?

The first interpretation seems to refer to something about the metaphysics of the object itself, while the second pertains to the relationship between an object and an observer.\cite{Szangolies,Brukner}

The transition from inequalities to non-locality can be schematised:
\begin{enumerate}
	\item\label{v1} QM produces probability values impossible in classical probability.
	\item\label{v2} Therefore, there exist indefinite properties that cannot be assigned a priori.
	\begin{itemize}
		\item Alternatively: quantum objects inherently possess fuzzy properties.
	\end{itemize}
	\item\label{v3} Therefore, the correlation of entangled objects is mediated by the instantaneous exchange of information, in a magical process, at the moment of state vector collapse.
\end{enumerate}

\ref{v1} is solid, it is the central result of Bell's work. \ref{v2} is weak or at least ambiguous. \ref{v3} seems reasonable, given \ref{v2}, but requires a leap of faith in a magical phenomenon. Probably \ref{v3} is widely accepted because the collapse itself is a magical phenomenon. Therefore, \ref{v3} does nothing more than add a bit of harmless nonsense to an already broadly irrational scheme.

\ref{v2} and \ref{v3} have dominated and still dominate the debate on the foundations of QM. \ref{v2} is directly related to the ontology of quantum objects and \ref{v3} is psychologically striking due to the apparent violation of relativity and its magical nature.

However, \ref{v2} and \ref{v3} are informal deductions, and not very robust, derived from \ref{v1}. The answer to the opening question "What exactly is bizarre about QM?" is all in \ref{v1}. QM is bizarre in that it violates classical probability, QM violates WE.

Let's take a step back. Forget about non-locality and quantum ontology. The new question to investigate is "How is it possible to violate WE? And what can legitimately be deduced from it?".\cite{Poletti1}
\section{A Logic Problem.}

\begin{quote}
	How is it possible to violate the Wigner-D’Espagnat inequality?
\end{quote}

A leap has been made, from physics to logic. In this new domain, there are no complex ontological evaluations, nor are there complicated structures such as space-time in which to define a sophisticated concept like "locality".

As we have seen, the Wigner-D’Espagnat inequality is a basic inequality in set theory, the demonstration of which is equally elementary. We repeat it:

If an element $x \in (A, C)$, then either $x \in B$, in which case $x \in (A, B)$, or $x \in \neg B$, in which case $x \in (\neg B, C)$.

The proof of WE relies on the law of the excluded middle, and the strongest temptation to violate WE is to violate this fundamental principle\cite{intuitionistic}. It makes sense, the narrative of quantum mechanics is replete with propositions in this direction: "the electron is both here and there", "the electron is both a wave and a particle", and so on. But violating the law of the excluded middle comes at a tremendous cost, as almost every known theorem depends on it either directly or indirectly.

Moreover, logic does not deal with the things of the world, but defines the rules of the language with which we describe the world; it is not in itself, "violable".

Imagine that an alien lands on earth and tells us: "On our planet, all aliens are mortal, but one of these aliens, Socrates, is immortal". How would we react? Would we have discovered that reality violates syllogisms? Clearly not. Syllogisms are not a constraint on reality but on language; we would probably simply ask the alien: "What do you mean by 'all'?"

This is what logic does: it defines the correct use of certain abstract terms such as "is", "exists", "all", "none". A real event that contradicts logic cannot exist by construction\cite{Tractatus}.

In the specific case, $\neg p$ indicates the only possible alternative to $p$. Attempting to violate these principles, or trying slightly more subtle approaches such as violating double negation, appears to be a contrivance in any case. Attempts in this direction, from Peirce\cite{Peirce} to fuzzy logic\cite{Zadeh}, are not viable.

Additionally, by adopting philosophical orientations closer to logical empiricism\cite{logicalempiricism}, the Platonic approach à la Penrose\cite{Penrose}, or the radical mathematical universes of Tegmark\cite{tegmark}, that is, by assigning greater ontological weight to logic, the argument can only become more robust. In these cases, logic becomes even more inviolable.

To get out of the mire, the only promising route seems to be observing that in logic too, there is a subtle problem of observer and object of observation. Consider a proposition $p$. The proposition $p$ is not true or false in itself, but is true or false depending on the formal system of axioms chosen. In a system of axioms that includes, for example, $p\land q$, $p$ will be true, in a system of axioms that includes $\neg p\land q$, it will be false. The truth or falsity of $p$ depends on the context.

Not only that, but in the context where $p$ assumes a definite truth value, a stronger logical condition applies: not only is $p$ true (or false), but it is also \textit{decidable}.

We identify a system of axioms with an observer $O$, an abstract entity that has the information of the axioms with which some theorems can be inferred. Relative to $O$, the proposition $p$ could be:
\begin{itemize}
	\item \textit{Manifestly true}: that is, $p$ is provable.
	\item \textit{Manifestly false}: that is, $\neg p$ is provable.
	\item \textit{Undecidable}: that is, neither $p$ nor $\neg p$ is provable.
\end{itemize}
The subtle difference between truth and provability, the core of G\"{o}del's theorems\cite{Godel}, will be the key to "violating" WE, the key to making sense of quantum superpositions of the type $\ket{yes}+\ket{no}$. As mentioned, WE cannot really be violated without altering classical logic. Consider its stronger version
\begin{equation*}
	\overline{(A,B)}\cup\overline{(\neg B,C)}\supseteq\overline{(A,C)}
\end{equation*}

The things that are \textit{manifestly} A and B, united with the things that are \textit{manifestly} not B and C, include the things that are \textit{manifestly} A and C.

In this strong form, WE is immediately violated by an element $x$, manifestly $A$ and $C$ and for which, relative to $O$, the property $B$ is undecidable.

What exactly is happening? It happens that WE is true; it's a valid theorem of set theory. However, concretely, a real observer $O$ (a system of axioms) "measuring" this theorem would invariably be in the stronger condition of having to verify not only its truth but also its provability in $O$ and thus would face a deadlock in the presence of any undecidable properties.

More generally, a set $A$ can be thought of, in absolute terms, as well-defined by entities $x$ that belong to the set $A$. However, relative to a concrete observer $O$, $A$ is only defined for those elements whose membership in $A$ is true and provable, leaving a margin of unknowability on a residue of possible undecidable elements.

Observe that there is no violation of classical logic. The ternary system (manifestly true, manifestly false, undecidable) is not an alternative to the binary system (true, false), but is simply the inclusion of the concept of provability in the discussion.

Classical probability is defined as the \textit{ratio of the number of favourable events to the number of possible events}. In set terms:
\begin{equation*}
	|p|=\frac{||P||}{||P||+||\neg P||}=\frac{||P||}{||P||+||P^c||}
\end{equation*}

But relative to $O$, the measure of a set can be concretely undecidable.

Probability straddles the mathematical formalism and the pragmatics of measurement.
However, the classical definition has a metaphysical flavour, and relative to a concrete, real observer $O$, it may not be applicable.

A very similar situation arises with Boolean algebra. It is certainly correct. Assigning the value 1 to true propositions and the value 0 to false ones, the logical combination of propositions can be expressed as products, sums, and differences of those numbers. But there's a problem: if $p$ is undecidable, Boolean algebra ceases to be concretely applicable by a certain observer for whom $p$ is undecidable.

The logic problem leads to a probability problem: how to define probabilities in the case of undecidable properties?

\section{A Probability Problem}
\begin{quote}
	How to construct a theory of probability that incorporates undecidable propositions?
\end{quote}
This question has been formally treated in \cite{Poletti2}, here only a brief summary will be provided.

The most natural definition of probability in relation to an observer $O$ is
\begin{equation}\label{qprob}
	[p]=\frac{||\overline P||}{||\overline P||+||\overline{\neg P}||}
\end{equation}

That is, \textit{the ratio between the number of manifestly valid events and the number of manifest events}.

We will call "quantum probability" the theory of probability that derives from \ref{qprob}.

Joint probability in classical probability is dominated by Bayes' theorem, so:
\begin{equation*}
	|p||q|_p=|q||p|_q
\end{equation*}

Where $|q|_p$ and $|p|_q$ are "the probability of $q$, given $p$" and "the probability of $p$, given $q$".

Conversely, from \ref{qprob} it is easy to obtain:

\begin{equation*}
	[p][q]_p\neq [q][p]_q
\end{equation*}

Quantum probability is intrinsically non-commutative. This makes sense, measuring $p$ implies an upheaval on the possibility of measuring $q$ as the possibility of verifying $q$ while keeping $p$ undecided decays. To construct an adequate algebra for quantum probability, it is appropriate to associate to $p$ and $q$ Hermitian linear operators (projectors) $\textbf{P}$ and $\textbf{Q}$ that act on a certain state vector $\ket{\psi}$. In a few steps, the Born rule is obtained:
\begin{equation*}
	[p]=\expval{\bm P}{\psi}
\end{equation*}

Moreover, the necessity of complex numbers in the algebra is directly connected to the necessity to respect the law of excluded middle while maintaining an additional degree of freedom to define undecidable states. The quantum probability defined by \ref{qprob} eventually coincides with the first postulates of QM.

The relationship between classical and quantum probability is given by 

\begin{equation*}
	[p]=|p|_{\bar{p}}
\end{equation*}

meaning, the quantum probability of $p$ is equal to the classical probability of $p$, given "$p$ is decidable".

QM has nothing illogical but abides, without exceptions, by classical logic. WE, in particular, is not exactly violated. What is violated is a strong version of WE which is the only version practically feasible by a concrete observer $O$.

In this sense, QM is a relational theory of measurement\cite{Rovelli}. QM does not dictate how a system $S$ is in itself, but how an observer $O$ interacts with $S$. The state vector $\ket{\psi}$ does not define an ontology of reality but at most an ontology of information relative to $S$ available in $O$.

Does this mean that it is possible, at least in principle, to elaborate a complete theory, albeit not practically feasible, a hidden variable theory beyond QM?

And this again shifts the focus to the philosophy of science.

\section{A Problem in Philosophy of Science.}

\begin{quote}
	Is it possible to complete QM?
\end{quote}

It is certainly possible to imagine constructing such a theory, but it cannot be a physical theory.

Physics is rigidly subjected to the scientific method and is, above all, a theory of measurement and only secondarily, potentially, a theory of reality. What every rational person implicitly accepts is that the Galilean method is the guide towards reality. That is, what science does is to assume that a good theory of reality should produce good experimental predictions.

This is sensible but there is a problem, the theory of measurement is always by construction relative to the observer who carries out that measurement. Even admitting that the "true" theory of probability is classical probability, physics cannot make use of it as it is forced by the scientific method to predict the concrete measures of observer $O$ and not theoretical metaphysical values. Even if we remove from reality any form of "quantum ontology", physics remains quantum physics, a theory of measurement in $O$.

So, has physics come to an end? Or is it time to abandon the scientific method? Certainly not!

It is time to deeply assimilate the fact that the greatest strength of physics, the scientific method, is an extraordinary wealth but also a constraint that prevents physics from being directly a theory of reality (metaphysics\cite{Poletti3}), and always, first and foremost, a theory of measurement.

This is what quantum mechanics is, an extraordinary theory of measurement, a theory that illustrates how $O$ describes and interacts with a system $S$.

In practical terms, this appears to suggest that any completion of Quantum Mechanics entails a metaphysical slippage. Bohm, in this respect, supplements QM with hidden variables that must be profoundly so, not truly assignable as initial conditions in a real experiment. In this theory, the proposition "the particle has coordinates $(x,y,z)$" is, therefore, metaphysical.

From Bell's work, it is not permissible to deduce non-locality, but something much weaker and significant. In a certain system $O$, there are propositions relative to reality, to the universe, that are undecidable. That is, it is not true that in every small arbitrary spatiotemporal vicinity, the entire information of the whole universe and its history is available.

This fact can also be considered obvious. However, it is notable how classical physics violates this basic principle.

A Newtonian particle of mass $m$ communicates its presence everywhere in the universe through a holomorphic function, its gravitational field, which allows every other body in the universe to deduce exactly its position and velocity. This is how Neptune and Pluto were discovered.

This is what QM teaches us, the boundary conditions are not always sufficient to deduce the entire state of the system. The information is not available everywhere. Every observer experiences some undecidable properties.

\section{Conclusions}

The interpretation of QM discussed here is, in some respects, disappointing, perhaps frustrating, as it wipes away all the magic of quantum mechanics at one stroke, without replacing it with anything equally striking. In this perspective, it is entirely inappropriate to say that "an electron is both here and there" or that "an electron takes all possible paths simultaneously." Reality is simply reduced to an electron that will be somewhere and to our objective impossibility of assigning it a well-defined position.

Likewise, it is inappropriate to say that "a quantum computer performs many calculations in parallel", leaving the listener astounded, but, much more banally, quantum computing exploits the peculiarities of quantum probability to produce new efficient algorithms.

And similarly, the many worlds\cite{Everett, Vaidman} and the many minds\cite{Albert} are entirely superfluous, as is any role of consciousness or non-locality. Finally, "quantum logic" does not indicate anything well-defined, as QM adheres to the rules of the usual classical logic. Are we ready to accept that QM might be a theory, after all, "normal"?

Einstein was right, QM is incomplete, it does not describe all things in the world. The state vector is not even remotely a thing of the world, no more than any probability distribution in any statistical problem is.

But Bohr was also right, QM is not completable because it does not deal with the reality of the world, but only with what we can measure about the world.

Physics naturally aspires to say something about the world, something profoundly true about the world. But physics is subject by construction to the scientific method and cannot, properly, produce a theory of the world, but at most a theory of measurement, and the two do not coincide.

QM shows to all of us this limit, the limit where a really powerful theory in predicting measurements says little about what the world is. Physics is always, by construction, relational. It always deals with what an observer $O$ (ourselves in the first place) observes around him. And these observations, these measurements are subject to the apparent logical paradoxes of quantum mechanics for the sole fact that we forget to constantly remind ourselves that physics does not deal with the truth of generic propositions $p$, but with constrained propositions: "$p$ is demonstrable in $O$".

Classical probability and quantum probability differ, in my opinion, not because of extravagant properties of quantum objects, but because they address different problems. Consider the following scenarios:

\begin{itemize}
	\item Determining the probability that flipping a coin will result in heads.
	\item Determining the probability that a trick coin, which has heads or tails on both sides, will result in heads.
\end{itemize}

The first scenario is a problem of classical probability. The answer is certainly $1/2$ and can be well framed within classical, subjectivist, or frequentist perspectives, depending on one's philosophical preferences.

The second problem is more ambiguous and may not even be a problem of probability in a strictly frequentist view, but it is certainly a legitimate problem in physics. Trick coins exist, and it is legitimate to study their evolution and assess the possible outcomes of experiments.

We know that experiments on such a coin, from the second one onwards, will be entirely determined by the outcome of the first experiment (which is exactly what happens with measurements of spin or polarization). The problem is to quantify the outcome of this first experiment.

The classical definition of probability still applies, after all, without difficulty: there are, a priori, two possible outcomes, one of which is favorable. On the other hand, every consideration related to a genuine process of deterministic chaos falls away. The complex spinning of the coin during the flip does not entail any path in a phase space densely populated with possible heads or tails outcomes. In fact, the physical process of flipping the coin becomes irrelevant, the trick coin might as well not be flipped, but simply observed.

A physical theory that deals with problems like 2) will have a probabilistic nature different from classical probability, where the presence of probability is not triggered by a classic random process but by the impossibility, given the axioms, to demonstrate the conclusion. A problem of undecidability.

What kind of probability theory results from this? Does Bayes' theorem still hold? What epistemology does it entail? Does this theory violate the Wigner-D'espagnat inequality?

I hope to have managed to show, or at least to insinuate the doubt, that the resulting theory is QM, a theory of the probabilities of undecidable propositions, an algebra of the semantic relations between undecidable propositions.

\section{Acknowledgements}

The initial title of this work was "Bertlmann's Socks in a G\"{o}delian Perspective". However, the first reviewer noted that the arguments discussed here do not directly involve any specific G\"{o}del theorem. Furthermore, according to an unverified story\cite{WheelerGodel}, Wheeler was once in G\"{o}del's office discussing similar themes and was rudely dismissed.

Regardless of the truth in this story, it would be inappropriate to associate G\"{o}del's name with the positions described here without any evidence that G\"{o}del himself developed thoughts in this direction. The term "Viennese" more generically refers to the reflections of the Vienna Circle on logic and its role in the world, creating a pleasant rhetorical tension with Copenhagen and an equally rich source of reflections on QM. I believe it to be a more honest title and thank the (unknown) reviewer for this and for several other extremely accurate observations that have led to this second draft.
\nocite{*}

\bibliography{Eng}

\end{document}